# Quantum crystallographic charge density of urea

Michael E. Wall[1,*]

[1]Computer, Computational, and Statistical Sciences Division, Los Alamos National Laboratory, Los Alamos, New Mexico, USA

*Correspondence: mewall@lanl.gov

**Synopsis**

A charge density model of urea was calculated using quantum theory and was refined against publicly available ultra high-resolution X-ray diffraction data. The quantum model differs from a multipole model but agrees comparably with the data; quantum crystallography therefore can provide unique and accurate charge density models.

**Abstract**

Standard X-ray crystallography methods use free-atom models to calculate mean unit cell charge densities. Real molecules, however, have shared charge that is not captured accurately using free-atom models. To address this limitation, a charge density model of crystalline urea was calculated using high-level quantum theory and was refined against publicly available ultra high-resolution experimental Bragg data, including the effects of atomic displacement parameters. The resulting quantum crystallographic model was compared to models obtained using spherical atom or multipole methods. Despite using only the same number of free parameters as the spherical atom model, the agreement of the quantum model with the data is comparable to the multipole model. The static, theoretical crystalline charge density of the quantum model is distinct from the multipole model, indicating the quantum model provides substantially new information. Hydrogen thermal ellipsoids in the quantum model were very similar to those obtained using neutron crystallography, indicating that quantum crystallography can increase the accuracy of the X-ray crystallographic atomic displacement parameters. The results demonstrate the feasibility and benefits of integrating fully periodic quantum charge density calculations into ultra high-resolution X-ray crystallographic model building and refinement.

## Introduction

Efforts to increase the accuracy of charge density models from X-ray crystallography mainly have focused on fitting the Bragg data using functions that are more expressive than the usual free-atom spherical distributions. Stewart (Stewart, 1969) proposed using general scattering factors that are the products of atom-centered orbital wave functions, and restrictions to better match the number of free parameters to the number of reflections in fitting (Stewart, 1970). Coppens et al. (Coppens *et al.*, 1971) separated the free atom charge density into core and valence components, and allowed them to be centered on different positions. Dawson decomposed the charge into symmetric and antisymmetric components centered on each atom (Dawson, 1967a), and expanded each atom-centered charge density in spherical harmonics (Dawson, 1967b). Hirshfeld developed a least-squares method that models aspherical atomic charge densities using basis functions related to spherical harmonics, but with alternative symmetry properties (Hirshfeld, 1971). Spherical harmonic-related methods were integrated into multipole refinement computer programs that are used when charge density models are desired (Hansen & Coppens, 1978, Hirshfeld, 1977a, Craven & Weber, 1977, Stewart & Spackman, 1983, Jelsch *et al.*, 2005, Volkov *et al.*, 2006).

Although less well exploited than multipole methods, the potential for combining quantum theory and X-ray diffraction to obtain accurate charge density models of molecular crystals has been long appreciated (Lipscomb, 1972). This combination has been termed quantum crystallography (Massa *et al.*, 1995). The high computational cost of quantum electronic structure calculations has been a major barrier to exploiting the theory for crystallography; however, recent linear scaling methods have made calculations possible for large systems (Bowler & Miyazaki, 2010, 2012, Goedecker, 1999, VandeVondele *et al.*, 2012), and fast quantum molecular dynamics simulations for systems approaching $10^4$ atoms with $10^5$ time steps are now possible (Mniszewski *et al.*, 2015). Methods for using of quantum theory to calculate crystallographic charge density models for all but the largest systems therefore might soon be within reach, not only for small-molecule crystallography (Capelli *et al.*, 2014) but also for macromolecular crystallography.

Several methods have been proposed for quantum crystallography, including the method of kernel projector matrices (Massa *et al.*, 1995) and fitting of wave functions to diffraction data (Jayatilaka, 1998). One method that is showing promise in practical applications is Hirshfeld Atom Refinement (HAR) (Bruning & Feil, 1992, Capelli *et al.*, 2014, Jayatilaka & Dittrich, 2008). In HAR, the static charge density of a molecule is calculated using quantum theory and is partitioned into individual atom contributions using Hirshfeld's stockholder method (Hirshfeld, 1977b). The partitioned charge is used to calculate aspherical atomic structure factors that are substituted for the usual structure factors in crystallographic refinement, considering both the atomic positions and displacement parameters (Bruning & Feil, 1992). Whereas Bruning and Feil (Bruning & Feil, 1992) originally decomposed the charge density into individual atom contributions using a multipole expansion; the more recent implementation of Jayatilaka & Dittrich and coworkers (Capelli *et al.*, 2014, Jayatilaka & Dittrich, 2008) directly makes use of a Becke grid for individual atom charge densities. The HAR method has been automated to apply iterative updates of the quantum electronic structure calculation during refinement of atomic positions (Capelli *et al.*, 2014). So far HAR has been limited to gas phase electronic structure calculations, with cluster charges placed at symmetry-related positions to approximate the crystal environment.

Whether HAR or other quantum crystallography methods will be adopted widely depends critically on whether they will substantially increase the accuracy of X-ray crystallography models. To date the main focus on the accuracy of HAR has been whether it yields molecular geometry and atomic displacement parameters that are consistent with neutron crystallography. The results here have been promising: applications to X-ray diffraction from crystalline benzene and urea (Jayatilaka & Dittrich, 2008), L-Phenylalaninium hydrogen maleate (Woinska *et al.*, 2014) , and a Gly-L-Ala dipeptide (Capelli *et al.*, 2014) found that HAR bond distances agreed very well with neutron crystal structures, overcoming known deficiencies in spherical atom charge density models (Lipscomb, 1972) . Atomic displacement parameters from HAR similarly agreed reasonably well with the neutron crystal structures.

This study addresses a major factor that so far has been lacking in evaluating quantum crystallographic methods: the accuracy of the charge density model. Here, charge density models for crystalline urea are obtained using spherical atom, atomic multipole, or quantum methods. For the quantum method, the HAR method (Jayatilaka & Dittrich, 2008) is adapted for crystalline phase electronic structure calculations performed using VASP (Kresse & Furthmuller, 1996). Electronic structure calculations using VASP previously were performed on hexachlorobenzene for comparisons to the X-ray crystallographic multipole charge density (Aubert *et al.*, 2011), but without allowing for individual ADPs. The novel aspect of the present method therefore is the combination of a crystalline phase density-functional-theory-based electronic structure calculation with an atomic displacement model from HAR. The results indicate that HAR can yield not only molecular geometries and ADPs that are similar to the neutron crystal structure, but also both 2Fo-Fc maps and static charge densities that are distinct from the multipole model, but that nevertheless agree comparably with the experimental data. Quantum crystallography therefore can yield accurate charge densities that are consistent simultaneously with theory and experiment.

**Methods**

*Diffraction data and initial crystal structure.* Ultra high-resolution urea synchrotron diffraction data were obtained from (Birkedal *et al.*, 2004) at URL <http://journals.iucr.org/a/issues/2004/05/00/xc5013/xc5013Isup7.hkl>. These data were collected at a temperature of 123 K using a wavelength of 0.5996(1) Å, and were merged into 1045 unique reflections (992 positively valued) extending to 0.347 Å resolution. The data were consistent with a P-$42_1m$ unit cell (space group

113), with *a*=*b*=5.5780(6) Å, *c*=4.6860(7) Å, α=β=γ=90°. Other data collection details are published in (Birkedal *et al.*, 2004), Table 1. The multipole refined urea crystal structure was obtained from (Birkedal *et al.*, 2004) at URL [http://journals.iucr.org/a/issues/2004/05/00/xc5013/xc5013sup1.cif](http://journals.iucr.org/a/issues/2004/05/00/xc5013/xc5013sup1.cif). The hydrogen parameters of this model were copied from a 123 K neutron crystal structure (Swaminathan *et al.*, 1984).

*Spherical atom and multipole crystallographic models*. The program SHELXL (Sheldrick, 2008, 2015) version 2014/1 was used to refine a spherical atom model of urea. Atomic coordinates and anisotropic atomic displacement parameters (ADPs) were refined for all atoms, in addition to an overall scale factor (27 parameters in all). The experimental temperature of 123 K was selected for geometry restraints. SHELXL reported agreement factors for the final model are: $R1=0.0370$, $wR2=0.0796$, and SHELX goodness of fit = 0.639 for all reflections. Mean unit cell charge density maps Fo, 2Fo-Fc, and Fo-Fc were calculated using the program *shelx2map* provided in the SHELX software distribution, using the refined .fcf file as the input, with default weighting, yielding a map of dimensions 56x15x42 for the asymmetric unit. The maps were expanded to P1 using CCP4 (Stein *et al.*, 1994) *mapmask* and were interpolated to a 64x64x64 grid using CCP4 *maprot.*

The program MoPro (Guillot *et al.*, 2001, Jelsch *et al.*, 2005) version 14.06 was used to refine a multipole charge density model of urea. Refinement was based on the ultra high-resolution data and structure from (Birkedal *et al.*, 2004). 20 cycles of automated density refinement were performed using the REFI DENS method. The total number of free parameters was 37: one scale factor; five valence (VAL); five $\kappa_1$ (K1); five $\kappa_2$ (K2); and 21 $P_{LM}$ (PLM) multipole parameters. The atom coordinates and ADPs were kept constant. MoPro reported agreement factors for the final model are (in %): RF=2.416, wR2F=1.068 RI=2.122, wR2I=2.117, and gof_F=2.293 for 992 nonzero reflections. Charge density maps were calculated using the MoPro supplied program VMoPro. The Fo, 2Fo-Fc, and Fo-Fc maps were computed by Fourier reconstruction using the FOUR method using the refined .par file and .FOUR file as inputs, with default resolution limits and the FFT method, yielding maps on a 92x92x80 grid; these maps were interpolated to a 64x64x64 grid using CCP4 *maprot* (Stein *et al.*, 1994). The total static crystalline charge density was computed using the VMoPro STAT method, with a 10 Å selection for grid limits, grid cube dimensions in fractional coordinates, the origin at (0,0,0), a maximum coordinate value of 0.9844 in each dimension, a 10 Å margin around the grid for contributing atoms, and 64x64x64 grid points. MoPro charge densities were scaled to yield a total charge of 64 electrons in the unit cell.

*Quantum crystallographic models*. A custom implementation of the original Hirshfeld atom refinement method (Jayatilaka & Dittrich, 2008) was used to obtain quantum crystallographic models. Quantum charge density calculations were performed using atomic coordinates from each of three different models: the SHELX refined spherical atom structure; the neutron crystal structure of (Swaminathan *et al.*, 1984); and the multipole model of (Birkedal *et al.*, 2004). An expanded unit cell with 16 atoms was generated using the Computational Crystallography Toolbox (cctbx) (Grosse-Kunstleve *et al.*, 2002) by applying P-42$_1$m symmetry to the 5-atom asymmetric unit. *Ab initio* density functional theory calculations were performed using VASP (Kresse & Furthmuller, 1996) version 5.3.3. Instead of pseudopotentials, the PAW method was used, with PAW_PBE parameters (Kresse & Joubert, 1999). The electronic structure was computed using 4$^3$=64 Monkhorst-Pack *k* points. Partial occupancies were calculated using Fermi smearing at the experimental temperature of 123 K. As there are fewer than 20 atoms in the expanded urea unit cell, LREAL=.FALSE. was used to evaluate projection operators in reciprocal space, as recommended in the VASP documentation. The valence charge density $v(x)$ was calculated for the expanded P1 unit cell. In addition, using the same VASP PAW method as for the molecular calculation, 16 crystalline core charge densities $c_i(x)$ and 16 crystalline free-atom ("promolecule") charge densities $f_i(x)$ were obtained for each individual atom *i*.

To achieve the desired model accuracy, all VASP charge densities were calculated on a 128x128x128 grid spanning the unit cell. For crystallographic refinement, the densities were decimated to a 64x64x64 grid.

X-ray structure factors were calculated using both new and existing tools in Lunus software (Wall, 2009), which was originally designed for analysis and modeling of diffuse X-ray scattering data (Wall, Clarage*, et al.*, 1997, Wall, Ealick*, et al.*, 1997, Wall *et al.*, 2014). The effect of ADPs was modeled using a Stockholder method

(Bruning & Feil, 1992, Hirshfeld, 1977b). The total valence density $v(x)$ was partitioned into atomic contributions using the equation

$$v_i(x) = q_i(x)v(x). \qquad (1)$$

Hirshfeld partitioning (Hirshfeld, 1977b) was used with weights $q_i(x)$ defined using the free-atom charge density $f_i(x)$,

$$q_i(x) = \frac{f_i(x)}{\sum_i f_i(x)}. \qquad (2)$$

Similar to (Bruning & Feil, 1992), ADPs were modeled by treating each partitioned atom charge density $a_i(x) = c_i(x) + v_i(x)$ as a rigid distribution, displaced along with the atom. However, in contrast with (Bruning & Feil, 1992), instead of using a multipole expansion, the charge density $a_i(uvw)$ was sampled on a rectilinear grid spanning the unit cell, indexed by $uvw$. This method is similar to that of (Jayatilaka & Dittrich, 2008), who used a radial-angular Becke grid for sampling. Here a rectilinear grid is chosen, as it corresponds precisely both to the VASP results and to the discrete sampling by the Bragg peaks in the crystallographic experiment. The partitioned atom structure factors were defined as $A_i(hkl) = DFT[a_i(uvw)]$, where DFT denotes a discrete Fourier transform. The DFT was computed using a fast Fourier transform (FFT) algorithm (Press *et al.*, 1999).

In the original Hirshfeld refinement method (Jayatilaka & Dittrich, 2008), the value $A'_i(hkl)$ of $A_i(hkl)$ after a coordinate shift $x'$ was obtained by multiplying $A_i(hkl)$ by a phase factor. Although multiplication by a phase factor is appropriate for arbitrary translations of a continuous distribution or an atom-centered grid, it is not appropriate for translations by fractional grid points on a fixed rectilinear grid such as is used here. The correct transformation instead requires a resampling of the shifted distribution $a_i(uvw)$ on the original grid (Appendix). The structure factors are obtained by transforming $x'$ to the grid coordinates $u'v'w'$, decomposing these coordinates into integer $(u_0 v_0 w_0)$ and fractional $(u_1 v_1 w_1)$ parts such that $0 \leq u_1 < 1$, $0 \leq v_1 < 1$, and $0 \leq w_1 < 1$, and using the following equation to calculate $A'_i(hkl)$,

$$A'_i(hkl) = A_i(hkl) e^{-\frac{2\pi i h u_0}{N_1}} e^{-\frac{2\pi i k v_0}{N_2}} e^{-\frac{2\pi i l w_0}{N_3}}$$

$$\times \left[ (1-u_1)(1-v_1)(1-w_1) + e^{-\frac{2\pi i h}{N_1}} u_1 (1-v_1)(1-w_1) \right.$$

$$+ e^{-\frac{2\pi i k}{N_2}} (1-u_1) v_1 (1-w_1) + e^{-\frac{2\pi i l}{N_3}} (1-u_1)(1-v_1) w_1$$

$$+ e^{-\frac{2\pi i h}{N_1}} e^{-\frac{2\pi i k}{N_2}} u_1 v_1 (1-w_1) + e^{-\frac{2\pi i h}{N_1}} e^{-\frac{2\pi i l}{N_3}} u_1 (1-v_1) w_1$$

$$\left. + e^{-\frac{2\pi i k}{N_2}} e^{-\frac{2\pi i l}{N_3}} (1-u_1) v_1 w_1 + e^{-\frac{2\pi i h}{N_1}} e^{-\frac{2\pi i k}{N_2}} e^{-\frac{2\pi i l}{N_3}} u_1 v_1 w_1 \right]. \quad (3)$$

The unit cell structure factor $F_c(hkl)$ was then calculated as

$$F_c(hkl) = \sum_i A'_i(hkl) e^{-2\pi^2 s_{hkl} \cdot U_i \cdot s_{hkl}}, \quad (4)$$

where $e^{-2\pi^2 s_{hkl} \cdot U_i \cdot s_{hkl}}$ is the Debye-Waller factor for the matrix $U_i$ of ADPs for atom $i$, and $s_{hkl}$ is the scattering vector corresponding to Miller indices $hkl$.

*Quantum model refinement.* Quantum refinements were performed starting with the spherical atom (S), multipole (M) (Birkedal *et al.*, 2004), and neutron crystallography (N) (Swaminathan *et al.*, 1984) atomic coordinates and ADPs. Model refinement was performed by minimizing the goodness of fit (GooF) statistic:

$$GooF = \left[\frac{\sum_{hkl}\left(I_o(hkl) - I_c(hkl)\right)^2 / \sigma_I^2(hkl)}{NDF}\right]^{1/2}, \quad (5)$$

where $I_c(hkl) = |F_c(hkl)|^2$, $I_o(hkl)$ and $\sigma_I(hkl)$ are the values and errors of the observed intensities, and the number of degrees of freedom *NDF* = 965 is the number of data points (= 992 non-negative intensity values in the merged data set), minus the number of free parameters in the fit (= 27, see below). A value of the GooF for each set of coordinates and ADPs was obtained by minimizing it with respect to an arbitrary scale factor between the calculated and observed reflection amplitudes. Each matrix $U_i$ was decomposed into eigenvalues and eigenvectors, and three Euler angles were computed from the eigenvectors, to obtain a set of independent parameters for efficient optimization. Optimization with respect to atom positions and eigenvalues and Euler angles from *U* matrices was performed in python using the *scipy.optimize.minimize* Powell method, using default settings. Due to the use of the Powell method, error bars were not obtained for the fitted parameter values. Eigenvalues were constrained to be positive. Symmetry of the atomic coordinates and ADPs was enforced explicitly using the following equations: X=0 and U23=0 for C, O atoms; Y=X+0.5 for all atoms; and U22=U11 and U13=U23 for all atoms. Enforcement of symmetry reduced the number of free parameters from 9 to 4 for the C, O atoms and to 6 for the N, H1, and H2 atoms. There were a total of 27 free parameters in the refinement, including the scale factor between the data and model (the same number as for spherical atom refinement, but without geometry restraints).

The mean unit cell charge densities ρ were calculated using Fourier reconstruction as $\rho(uvw) = DFT[F(hkl)]$. The experimental Fo, 2Fo-Fc and Fo-Fc maps were calculated by applying the model phases to the observations. Values of $|F_c(hkl)|$ were used in place of missing values of $|F_o(hkl)|$. Only complete grids were used for FFT calculations on the quantum models; reflections were not truncated using a resolution cutoff.

*Agreement factors.* The agreement of all models with the diffraction data was assessed using several standard statistics: GooF, wR2F, wR2I, RF, and RSR. The GooF (Eq. (5)) was used as the refinement target. The weighted R-squared factor for amplitudes, wR2F, was calculated as

$$wR2F = \left[\frac{\sum_{hkl}(|F_o(hkl)|-|F_c(hkl)|)^2/\sigma_F^2(hkl)}{\sum_{hkl}|F_o(hkl)|^2/\sigma_F^2(hkl)}\right]^{1/2}, \qquad (6)$$

where $|F_o(hkl)|$ and $\sigma_F(hkl)$ are the experimental reflection amplitudes and errors, and $F_c(hkl)$ is calculated using Eq. (4). The weighted R-squared factor for intensities, wR2I, was calculated as

$$wR2I = \left[\frac{\sum_{hkl}(I_o(hkl)-I_c(hkl))^2/\sigma_I^2(hkl)}{\sum_{hkl}I_o(hkl)/\sigma_I^2(hkl)}\right]^{1/2}; \tag{7}$$

the R-factor for amplitudes, RF, was calculated as

$$RF = \frac{\sum_{hkl}||F_o(hkl)|-|F_c(hkl)||}{\sum_{hkl}|F_o(hkl)|}; \tag{8}$$

the real-space R-factor, RSR, was calculated as

$$RSR = \frac{\sum_{uvw}|\rho_o(uvw)-\rho_c(uvw)|}{\sum_{uvw}|\rho_o(uvw)+\rho_c(uvw)|}, \tag{9}$$

where $\rho_o$ is the experimental Fo map. Calculated and observed values were scaled to minimize the RMSD prior to using Eqs. (5)-(9), and both $\rho_o$ and $\rho_c$ were offset to have zero mean prior to using Eq. (9). To enable fair comparison, all agreement factors were calculated using Lunus software tools. Values reported in primary references were very similar to those computed using Lunus.

## Results

The agreement factors for all quantum crystallographic models are the same (in % units): wR2F (target) = 0.9, wR2I = 1.8, RF = 2.3, RSR = 2.1, and GooF = 1.7 (Table 1). These are slightly better than the multipole model, which has values 0.2-0.3% higher for each. The quantum and multipole models agree much better with the data than the spherical atom model (Table 1).

Three-dimensional visualizations of the 2Fo-Fc and Fo-Fc maps for the spherical atom, quantum-M, and multipole models are shown in Fig. 1. The spherical atom and multipole 2Fo-Fc maps appear to be more similar to each other than they are to the quantum model. This appearance is supported quantitatively using a RSR statistic calculated between each pair of 2Fo-Fc real-space maps, using an appropriately modified Eq. (9). A value of 4.9% was obtained between the spherical atom and multipole models. By comparison, the RSR values between the quantum model and either the spherical atom (8.2%) or the multipole model (7.4%) were much greater. These values are all higher than the RSR of any of the models with the data (Table 1).

Visualization of contours in a 2D section including the C=O bond reveals that the quantum and multipole 2Fo-Fc maps are very different (Fig. 2A-2B). (Much of this difference might be an artifact in the multipole map calculation, as mentioned below.) Compared to the multipole model (Fig. 2B), the quantum-M model is smoother (Fig. 2A). The multipole model shows ripples surrounding core atoms and peaks away from atoms, including between bonded heavy atoms. The quantum-M model has some peaks away from the atom cores (Fig. 2A), but these are lower in magnitude compared to the multipole model. The quantum-M and multipole model Fo-Fc difference maps are broadly similar (Figs. 2C, 2D), with the larger deviations from the data in the multipole model along the C=O axis, consistent with the slightly higher values of agreement factors for this model (Table 1).

To investigate further the differences between the 2Fo-Fc maps of the quantum-M and multipole models, we compared the static total charge densities calculated using either VASP or MoPro. Both charge densities correspond to the multipole geometry (Birkedal *et al.*, 2004). There are visible differences (Figs. 3A, 3B); however, the differences are much smaller than in the 2Fo-Fc maps (Figs. 2A, 2B), and they coincide with atoms and bonds. The comparison suggests that the ripples in the 2Fo-Fc map from the multipole model are an artifact of the FOUR method implementation in VMoPro (e.g., a truncation of reflections beyond 0.347 Å resolution).

Subtracting the static charge densities from VASP and MoPro reveals substantial differences in the charge distribution along the C=O bond (Fig. 3C). These differences show a similar pattern of peaks and troughs as in the Fo-Fc map for the multipole model (Fig. 2D); by comparison, the Fo-Fc map of the quantum-M model shows smaller differences along the C=O bond (Fig. 2C). Combined, Figs. 2 and 3 indicate that the multipole static charge density contains deviations from the data in the C=O bond that are decreased in the quantum-M model.

The VASP and MoPro calculations were further compared using a Bader analysis of the net atom charges (Tang *et al.*, 2009) (Table 2). The theoretical VASP charges are similar for the spherical atom, multipole, and neutron structures. The main difference between these and the multipole model charges is for the C atom, which has a value of 4.06-4.08 electrons from the theoretical density, and 4.57 electrons from the multipole model density. This substantial 0.5 electron difference is compensated by smaller differences in the charges on the other atoms, which are between 0.06-0.09 electrons smaller in the multipole model.

The atomic coordinates of the quantum-M, quantum-N, neutron, and multipole models are all very similar (Table 3). The differences between these models and either the spherical atom or quantum-S model are small for the heavy atoms, but are larger for the hydrogens. The differences lead to a substantial deviation in the N-H1 bond for the spherical atom and quantum-S structure compared to the neutron structure (Table 4): the bond length is 1.006 Å in the neutron structure compared to 0.911 Å in the spherical atom and 0.810 Å in the quantum-S structure. The differences also lead to decreases in the C-N-H1 and C-N-H2 bond angles for both the spherical atom and quantum-S structures compared to the neutron structure (Table 5): the angles in the spherical atom model are about 2 degrees smaller, and the angles in the quantum-S model are about 4 degrees smaller than in the neutron structure. There is a corresponding increase in the H1-N-H2 angle for each compared to the neutron structure: 4 degrees for the spherical atom and 8 degrees for the quantum-S structure. The angle deviations for the quantum-S model are visible in the stick diagram in Fig. 3A; the effect is smaller but still perceptible for the spherical atom model (not shown).

The similarity of ADPs was assessed using the S statistic, which describes the deviation of the three-dimensional positional distribution of the atoms defined by the ADPs (Whitten & Spackman, 2006). The ADPs for the heavy atoms in the quantum models are very similar to the neutron models (Table 6; Fig. 4): the value of S for the C atom ranges from 0.04%-0.06%; the value for the O atom from 0.13% to 0.14%; and the value for the N atom from 0.27% to 0.32%. The similarities are comparable for the multipole model. The ADPs for the hydrogens in the quantum models are also similar to the neutron crystal structure, but to a lesser degree than the heavy atoms: the value of S for H1 varies from 1.61% to 2.25%; and the value for H2 varies from 2.34% to 3.46%. A high-level quantum theoretical calculation of vibrations of urea to obtain ADPs (Madsen *et al.*, 2013) yielded a comparable similarity for the heavy atoms (S = 0.12%, 0.16%, and 1.1% for C, O, and N, respectively) and a higher similarity for the hydrogens (S = 0.13% and 0.05% for H1 and H2, respectively). However, the similarities in (Madsen *et al.*, 2013) were computed after applying an overall scale factor with respect to the experimental ADPs; the similarities are considerably lower without applying the scale factor (S =

1.36%, 0.9%, 1.7%, 0.5%, and 0.65% for C, O, N, H1, and H2 respectively using the B3LYP/6-31G(d,p) method). The similarity of the spherical atom heavy atom ADPs to the neutron structure is high (S = 0.04%, 0.13%, and 0.21% for C, O, and N), and the similarity for the hydrogens is low, as expected for a spherical atom model (S = 25.65% and 5.91%). The multipole model hydrogen parameters were copied from the neutron structure and therefore are identical (Birkedal *et al.*, 2004).

To assess the convergence of the quantum refinement, as was done in previous HAR implementations (Capelli *et al.*, 2014, Jayatilaka & Dittrich, 2008) the electronic structure calculation was iteratively applied to each of the quantum models. In the iteration, the refined atomic coordinates were used to re-compute all charge densities using VASP, and the model was re-refined against the data using the new densities. The quantum-M and quantum-N models were essentially unchanged in the second iteration: the agreement factors remained the same as in Table 1; all of the fractional atomic coordinates changed by less than $5 \times 10^{-3}$, with maximal changes of $1 \times 10^{-3}$ for heavy atom coordinates; and the similarity statistic for the ADPs was 0.06% or lower for all atoms between the first and second iterations. In contrast, the quantum-S model showed divergent behavior: the agreement factors were slightly larger (by 0.1%-0.2%) for the second iteration; hydrogen fractional coordinates changed by as much as 0.05 (a 0.2 Å shift of the x- and y-position of the H1 atom); and the similarity statistic for the ADPs was as high as 2.6% (H1 atom), which is comparable to the value computed between the quantum models and the neutron model (Table 6).

**Discussion**

The agreement of the quantum crystallographic models of urea with ultra high-resolution data compares favorably to the multipole model. Both the 2Fo-Fc map and the total static charge density are substantially different between the quantum and multipole models, however. The differences in 2Fo-Fc appear to be due to an artifact in the multipole map, as they contain ripples that do not coincide with atom positions or bonds (Fig. 2A, B). The differences in the static charge density, however, appear to be real, with notable differences both in the electronic structure of the C=O bond (Fig. 3C) and in the 0.5-electron higher negative charge associated with the C atom for the multipole model (Table 2) . The difference in the C atom charge is consistent with the multipole charge density study of (Birkedal *et al.*, 2004), which reported a 0.7-0.8 electron larger negative charge for the C atom in the multipole model compared to theoretical charge density calculations.

Whereas (Birkedal *et al.*, 2004) concluded the difference between their multipole model of urea and the theoretical charge density was due to inaccuracies in the quantum electronic structure calculation, this study suggests that the difference might instead be due to inaccuracies in the multipole model. The quantum charge densities were obtained using quantum theory and are consistent with the experimental data. The multipole model, although also consistent with the experimental data, is more weakly tied to the underlying theory, and relies on the fitting of many parameters. The possibility of inaccuracies in the multipole model is supported by a controlled study using synthetic data (De Vries *et al.*, 2000) which found that the charge density of urea could not be determined uniquely using multipole refinement; however, this support is tempered by the fact that the synthetic data did not extend to a resolution as high as the data in (Birkedal *et al.*, 2004).

The present results indicate that it would be worthwhile investigating whether HAR might produce more reliable interaction density models than are currently obtained using multipole methods. Compared to multipole refinement, HAR uses fewer parameters and relies on quantum theory for the increased expressiveness needed to model the aspherical component of the charge density. In addition, in HAR, the same quantum electronic structure method used for the crystal phase calculation can be used to for the gas phase. Thus, whereas calculating the multipole interaction density involves subtracting two densities that were obtained using substantially different methods, the HAR interaction density can be obtained by subtracting densities that are more comparable.

The thermal ellipsoids in the quantum models are both quantitatively (Table 6) and qualitatively (Fig. 4) similar to the neutron crystal structure. This was even the case for the quantum-S model, despite the lack of convergence seen in a second iteration of refinement and deviations in the geometry with respect to the neutron model (Tables 4, 5). This finding is consistent with studies in which the neutron crystallographic temperature factors of urea (Jayatilaka & Dittrich, 2008) and other systems (Capelli *et al.*, 2014, Woinska *et al.*, 2014) were found to be reproduced reasonably well using HAR. In particular, the previous urea study (Jayatilaka & Dittrich, 2008) used BLYP density functional theory with a cc-pVTZ basis and surrounding charge clusters to mimic periodic boundary conditions, and used the same starting structure as the present quantum-M refinement (Birkedal *et al.*, 2004). The following ADP values were obtained for hydrogen atoms (Jayatilaka & Dittrich, 2008) ($U_{11}, U_{33}, U_{12}$, and $U_{13}$ in Å$^2$ units): (0.0550, 0.0170, -0.0350, 0) for H1, and (0.0450, 0.0260, -0.0190, -0.0020) for H2. These values are similar to those found here (Table 6); in addition, values for heavy atoms differed by less than 0.001 Å$^2$ compared to those found here. The similarity statistics S computed with respect to the ADPs from the quantum-M model are (in % units): 0.01, 0.03, 0.06, 0.26, and 0.08 for the C, N, O, H1, and H2 atoms, respectively. The similarity of the ADPs here with those in (Jayatilaka & Dittrich, 2008), in addition to the consistency of the quantum refined ADPs in this study using different starting structures, indicates that HAR can yield estimates of ADPs that are robust to differences in starting structures and DFT methods.

The results for the quantum-M and quantum-N models indicate the potential advantages of quantum crystallography for accurate charge density studies. The lack of convergence and geometry deviations of the quantum-S model, however, indicate that challenges remain for the general applicability of these methods. The deviations of the quantum-S model can be traced back to deviations in the spherical atom model: although the heavy atoms are consistent with the neutron structure, the deviation in the hydrogen positions is more substantial (Table 3), leading to corresponding deviations in geometry (Tables 4, 5). These deviations are increased rather than decreased in the quantum-S model, which prevents the quantum refinement from converging on successive iterations.

Because it is not currently feasible to obtain neutron crystal structures for all systems of interest, the generalization of the quantum crystallography methods developed here to routine X-ray crystallographic structure determination will require improved modeling of hydrogen positions in the starting structure. The successful application of iterative electronic structure calculations in HAR applications to ammonia and Gly–l –Ala using spherical atoms models like those used here as input structures (Capelli *et al.*, 2014) indicates that a spherical atom model is adequate for at least some molecular crystals. It would be interesting to determine whether iterative HAR using the implementation of (Capelli *et al.*, 2014) converges using the present spherical atom model of urea as an input (as mentioned above, the study of (Jayatilaka & Dittrich, 2008) made use of the same model as the quantum-M refinement here). It is possible that hydrogen positions in spherical atom models would be sufficiently improved using methods that leverage information in structure databases (Bak *et al.*, 2011, Bendeif & Jelsch, 2007, Dadda *et al.*, 2012, Dittrich *et al.*, 2005, Dittrich *et al.*, 2009), which can place hydrogens to within $O(10^{-2})$ Å of the positions in neutron crystal structures.

There are many ways HAR may be extended, targeting, e.g., more accurate models of structure variation than ADPs, and larger systems. For larger systems, it will be especially important to assess the applicability of fast, approximate quantum electronic structure calculations (Mniszewski *et al.*, 2015) to quantum crystallography. Increasing the speed of calculations would enable the wider adaptation of HAR by adapting existing small-molecule crystallography workflows, and would provide a complementary quantum crystallographic alternative to multipole refinement (Jelsch *et al.*, 2000) for obtaining high-resolution charge density models of molecular crystals, including macromolecular crystals.

**Acknowledgment**

This study was supported by the US Department of Energy under Contract DE-AC52-06NA25396 through the Laboratory-Directed Research and Development Program at Los Alamos National Laboratory.

**Appendix**

**Transformation of the structure factors upon translation of the charge density**

Here the transformation is illustrated using the one-dimensional case; the extension to three dimensions in Eq. (3) is straightforward. The structure factors are given by the discrete Fourier transform (DFT) of a periodic charge density $\rho(u)$ sampled at fixed grid points $u$, with $\rho(u + N) = \rho(u)$. Let the charge density $\rho(u)$ correspond to the following continuous step-wise distribution:

$$\rho(x) = \sum_{u=0}^{N-1} \rho(u)[\theta(x - u) - \theta(x - u - 1)], \qquad (A1)$$

where the Heaviside distribution $\theta(x) = 1$ for $x \geq 0$ and $\theta(x) = 0$ otherwise. Translation of $\rho(x)$ by a shift $\Delta x$ yields a new charge density

$$\rho_{\Delta x}(x) = \rho(x - \Delta x) = \sum_u \rho(u)[\theta(x - \Delta x - u) - \theta(x - \Delta x - u - 1)]. \quad (A2)$$

Discrete sampling of this shifted charge density on the original fixed grid at integer $u$ yields

$$\begin{aligned} \rho_{\Delta x}(u) &= \int_u^{u+1} dx \, \rho(x - \Delta x) \\ &= \int_u^{u+1} dx \sum_{u'} \rho(u')[\theta(x - \Delta x - u') - \theta(x - \Delta x - u' - 1)] \end{aligned} \qquad (A3)$$

Decomposing the shift $\Delta x = u_0 + u_1$ into an integer component $u_0$ plus a positive fractional component $0 \leq u_1 < 1$ yields

$$\rho_{\Delta x}(u) = \sum_{u'} \rho(u') \int_u^{u+1} dx \, [\theta(x - u_0 - u' - u_1) - \theta(x - u_0 - u' - u_1 - 1)] \quad (A4)$$

which is only nonzero for the terms $u' = u - u_0$ and $u' = u - u_0 - 1$. Performing the integrals for these terms yields

$$\rho_{\Delta x}(u) = \rho(u - u_0)(1 - u_1) + \rho(u - u_0 - 1)u_1 \qquad (A5)$$

Define the structure factors $A(h) = DFT[\rho(u)]$ and $A_{\Delta x}(h) = DFT[\rho_{\Delta x}(u)]$. $A_{\Delta x}(h)$ is obtained from $A(h)$ by by applying the DFT shift theorem to (A5),

$$A_{\Delta x}(h) = e^{-\frac{2\pi i h u_0}{N}} A(h) \left[ (1 - u_1) + e^{-\frac{2\pi i h}{N}} u_1 \right]. \qquad (A6)$$

which is the one-dimensional version of Eq. (3) in the text. Eq. (A6) is exact, and demonstrates that $A_{\Delta x}(h) \neq e^{-\frac{2\pi i h \Delta x}{N}} A(h)$ if $\Delta x$ includes a nonzero fractional shift $u_1$.

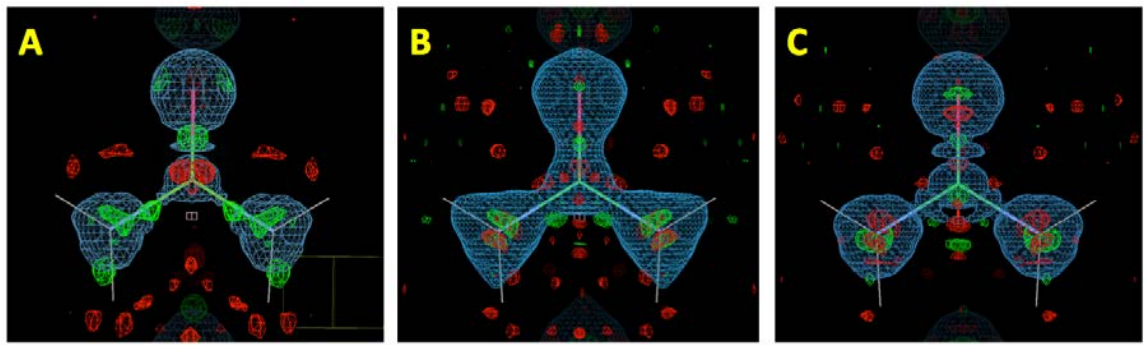

Figure 1. Comparison of 2Fo-Fc and Fo-Fc maps for (A) spherical atom, (B) Quantum-M, and (C) multipole models. Level charge density surfaces in 2Fo-Fc maps are rendered using a blue wireframe at a level of 1-sigma. Level surfaces in Fo-Fc maps are shown in green (positive e- density, negative charge) and red (negative e- density, positive charge) wireframes at 3-sigma. The figure was created using COOT (Emsley *et al.*, 2010).

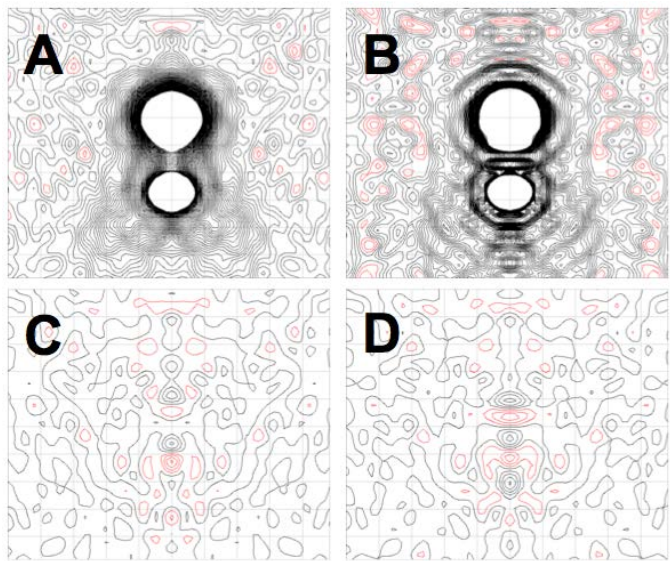

Figure 2. Comparison of 2D contours in the reconstructed mean unit cell charge density for the quantum-M (left), and multipole (right) models in the y=0 section. (A)-(B) 2Fo-Fc maps in 0.05 e- Å$^{-1}$ contours, to a maximum of 3 e- Å$^{-1}$. (C)-(D) Fo-Fc difference maps in 0.05 e- Å$^{-1}$ contours. Negative electron density contours in each panel are colored red. The view is along the same direction as that in Fig. 4, in the plane of the C=O bond. The orientation is such that x increases along the horizontal, and z increases along the vertical. The image was created using *mapslicer* in the CCP4 suite (CCP4, 1994).

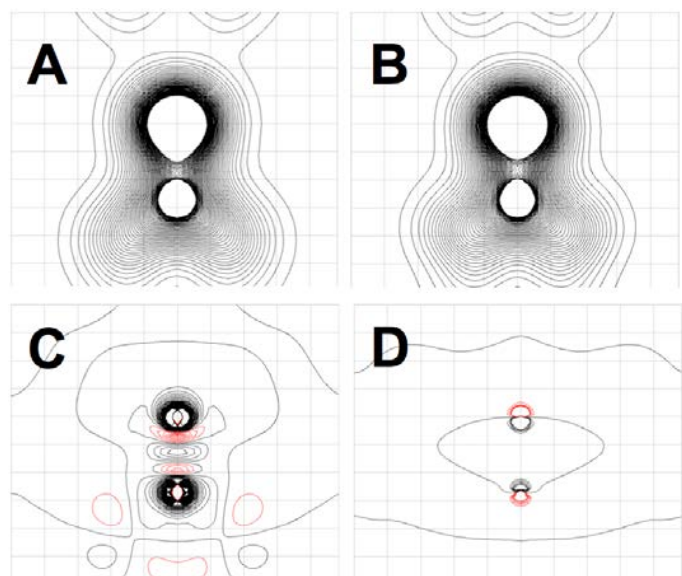

Figure 3. 2D contours in the static total charge densities derived from the multipole model atom coordinates (Birkedal *et al.*, 2004) (the section and orientation is the same as in Fig. 2). (A) Theoretical density computed using VASP. (B) Multipole charge density refinement in MoPro. (C) Density in (B) subtracted from density in (A). (D) Difference density computed by subtracting the theoretical density using quantum-M refined atom coordinates from the density in (A) (the view of the total density using the quantum-M structure is indistinguishable from (A)). Contours in all panels are in 0.05 e- Å$^{-1}$ intervals, to a maximum of 3 e- Å$^{-1}$. Negative electron density contours in panels (C) and (D) are colored red. The image was created using *mapslicer* in the CCP4 suite (CCP4, 1994).

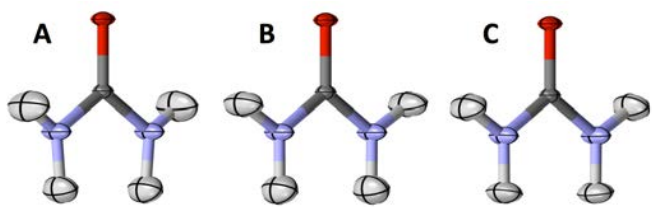

Figure 4. Thermal ellipsoids at 50% probability for (A) Quantum-S, (B) Quantum-M, and (C) neutron diffraction models. In each structure, the O atom is red, the C dark grey, the N blue, and the H1 and H2 light grey, with the H2 at the bottom of the molecule. The Quantum-N model is not shown as it is indistinguishable from the Quantum-M model; similarly, the multipole model is not shown as it is indistinguishable from the neutron. The latter is due to the fact that the hydrogen parameters of the multipole model were copied from the neutron model (Birkedal *et al.*, 2004). The image was created using Mercury (Macrae *et al.*, 2008).

Table 1. Values of crystallographic agreement factors for: spherical atom (Sphere); quantum starting with the spherical atom (Quant-S), Birkedal et al. (Birkedal *et al.*, 2004) multipole (Quant-M), and neutron (Quant-N) model; and multipole (Birkedal *et al.*, 2004) model. All values are in % units.

|          | wR2F | wR2I | RF  | RSR | GooF |
|----------|------|------|-----|-----|------|
| Sphere   | 3.4  | 6.8  | 3.8 | 4.2 | 6.5  |
| Quant-S  | 0.9  | 1.8  | 2.3 | 2.1 | 1.7  |
| Quant-M  | 0.9  | 1.8  | 2.3 | 2.1 | 1.7  |
| Quant-N  | 0.9  | 1.8  | 2.3 | 2.1 | 1.7  |
| Multipole| 1.1  | 2.1  | 2.6 | 2.3 | 2.0  |

Table 2. Atom charges based on Bader analysis of total static charge densities. VASP calculations correspond to the spherical atom (S), multipole (M), and neutron (N) structures. The MoPro calculation corresponds to the multipole (M) structure. Units are negative charge in electrons.

|          | C    | O    | N    | H1   | H2   |
|----------|------|------|------|------|------|
| VASP (S) | 4.08 | 9.26 | 8.33 | 0.51 | 0.50 |
| VASP (M) | 4.06 | 9.26 | 8.31 | 0.51 | 0.53 |
| VASP (N) | 4.06 | 9.27 | 8.31 | 0.51 | 0.53 |
| MoPro (M)| 4.57 | 9.20 | 8.22 | 0.46 | 0.44 |

Table 3. Coordinates for atoms in the asymmetric unit. Model labels in column 2 are as in Table 1, with the addition of the neutron model (Swaminathan *et al.*, 1984). Hydrogen model parameters for the multipole model were copied from the neutron model (Birkedal *et al.*, 2004) . Units are fraction of unit cell dimensions.

|   |   | X | Y | Z |
|---|---|---|---|---|
| C | Sphere | 0 | 0.5 | 0.3281 |
|   | Quant-S | 0 | 0.5 | 0.3279 |
|   | Quant-M | 0 | 0.5 | 0.3281 |
|   | Quant-N | 0 | 0.5 | 0.3277 |
|   | Neutron | 0 | 0.5 | 0.3280 |
|   | Multipole | 0 | 0.5 | 0.3282 |
| O | Sphere | 0 | 0.5 | 0.5964 |
|   | Quant-S | 0 | 0.5 | 0.5967 |
|   | Quant-M | 0 | 0.5 | 0.5966 |
|   | Quant-N | 0 | 0.5 | 0.5963 |
|   | Neutron | 0 | 0.5 | 0.5962 |
|   | Multipole | 0 | 0.5 | 0.5963 |
| N | Sphere | 0.1450 | 0.6450 | 0.1783 |
|   | Quant-S | 0.1452 | 0.6452 | 0.1782 |
|   | Quant-M | 0.1446 | 0.6446 | 0.1796 |
|   | Quant-N | 0.1447 | 0.6447 | 0.1785 |
|   | Neutron | 0.1447 | 0.6447 | 0.1785 |
|   | Multipole | 0.1447 | 0.6447 | 0.1790 |
| H1 | Sphere | 0.2438 | 0.7438 | 0.2793 |
|   | Quant-S | 0.2316 | 0.7316 | 0.2722 |
|   | Quant-M | 0.2571 | 0.7571 | 0.2837 |
|   | Quant-N | 0.2571 | 0.7571 | 0.2838 |
|   | Neutron | 0.2557 | 0.7557 | 0.2841 |
|   | Multipole | " | " | " |
| H2 | Sphere | 0.1382 | 0.6382 | -0.0363 |
|   | Quant-S | 0.1324 | 0.6324 | -0.0371 |
|   | Quant-M | 0.1432 | 0.6432 | -0.0343 |
|   | Quant-N | 0.1432 | 0.6432 | -0.0344 |
|   | Neutron | 0.1431 | 0.6431 | -0.0348 |
|   | Multipole | " | " | " |

Table 4. Bond lengths for alternative models. Model labels are as in Table 1. Units are Å.

|          | C=O   | N-O   | N-H1  | N-H2  |
|----------|-------|-------|-------|-------|
| Sphere   | 1.257 | 2.268 | 0.911 | 1.006 |
| Quant-S  | 1.259 | 2.271 | 0.810 | 1.013 |
| Quant-M  | 1.258 | 2.263 | 1.011 | 1.001 |
| Quant-N  | 1.258 | 2.266 | 1.012 | 0.996 |
| Neutron  | 1.257 | 2.266 | 1.006 | 1.000 |
| Multipole| 1.257 | 2.265 | 1.005 | 1.002 |

Table 5. Bond angles for alternative models. Model labels are as in Table 1. Units are degrees.

|  | O-C-N | N-C-N | C-N-H1 | C-N-H2 | H1-N-H2 |
|---|---|---|---|---|---|
| **Sphere** | 121.54 | 116.91 | 117.28 | 118.42 | 124.30 |
| **Quant-S** | 121.49 | 117.04 | 115.68 | 115.82 | 128.50 |
| **Quant-M** | 121.40 | 117.21 | 119.87 | 120.81 | 119.32 |
| **Quant-N** | 121.49 | 117.01 | 119.30 | 121.02 | 119.68 |
| **Neutron** | 121.54 | 116.92 | 118.99 | 120.82 | 120.20 |
| **Multipole** | 121.50 | 117.01 | 119.15 | 120.78 | 120.07 |

Table 6. Values of ADPs for atoms in the asymmetric unit. Labels in column 2 are as in Table 2. Units are Å². U11=U22 and U13=U23 by symmetry. The similarity statistic (S) with respect to the ADPs of the neutron model is computed following Ref. (Whitten & Spackman, 2006), in % units. Hydrogen model parameters for the multipole model were copied from the neutron model (Birkedal *et al.*, 2004) .

|   |   | U11 = U22 | U33 | U12 | U13 = U23 | S |
|---|---|---|---|---|---|---|
| C | Sphere | 0.0150 | 0.0070 | 0.0000 | 0.0000 | 0.04 |
|   | Quant-S | 0.0141 | 0.0061 | 0.0001 | 0.0000 | 0.05 |
|   | Quant-M | 0.0141 | 0.0061 | 0.0001 | 0.0000 | 0.04 |
|   | Quant-N | 0.0141 | 0.0060 | 0.0001 | 0.0000 | 0.06 |
|   | Neutron | 0.0147 | 0.0065 | 0.0001 | 0.0000 | 0.00 |
|   | Multipole | 0.0152 | 0.0068 | -0.0004 | 0.0000 | 0.03 |
| O | Sphere | 0.0199 | 0.0066 | 0.0020 | 0.0000 | 0.13 |
|   | Quant-S | 0.0194 | 0.0059 | 0.0019 | 0.0000 | 0.14 |
|   | Quant-M | 0.0194 | 0.0060 | 0.0020 | 0.0000 | 0.14 |
|   | Quant-N | 0.0194 | 0.0061 | 0.0020 | 0.0000 | 0.13 |
|   | Neutron | 0.0197 | 0.0063 | 0.0001 | 0.0000 | 0.00 |
|   | Multipole | 0.0196 | 0.0067 | 0.0016 | 0.0000 | 0.10 |
| N | Sphere | 0.0293 | 0.0096 | -0.0155 | 0.0001 | 0.21 |
|   | Quant-S | 0.0285 | 0.0087 | -0.0158 | 0.0000 | 0.32 |
|   | Quant-M | 0.0285 | 0.0085 | -0.0156 | 0.0001 | 0.29 |
|   | Quant-N | 0.0286 | 0.0087 | -0.0156 | 0.0001 | 0.27 |
|   | Neutron | 0.0286 | 0.0095 | -0.0147 | 0.0002 | 0.00 |
|   | Multipole | 0.0293 | 0.0096 | -0.0157 | 0.0000 | 0.23 |
| H1 | Sphere | 0.0295 | 0.0468 | -0.0127 | -0.0184 | 25.65 |
|   | Quant-S | 0.0550 | 0.0259 | -0.0392 | -0.0019 | 1.61 |
|   | Quant-M | 0.0495 | 0.0168 | -0.0295 | 0.0019 | 2.13 |
|   | Quant-N | 0.0490 | 0.0172 | -0.0296 | 0.0022 | 2.25 |
|   | Neutron | 0.0440 | 0.0216 | -0.0223 | -0.0031 | 0.00 |
|   | Multipole | " | " | " | " | " |
| H2 | Sphere | 0.0415 | 0.0206 | 0.0099 | -0.0024 | 5.91 |
|   | Quant-S | 0.0380 | 0.0227 | -0.0191 | 0.0015 | 2.34 |
|   | Quant-M | 0.0409 | 0.0270 | -0.0187 | -0.0013 | 3.43 |
|   | Quant-N | 0.0410 | 0.0272 | -0.0186 | -0.0012 | 3.46 |
|   | Neutron | 0.0430 | 0.0141 | -0.0159 | 0.0020 | 0.00 |
|   | Multipole | " | " | " | " | " |